\documentclass{ws-p8-50x6-00}
\usepackage{cite}
\begin{document}
\newcommand{\PO}{I\!\!P}
\newcommand{\xpom}{x_{\PO}}

\title{Introduction to Diffraction\\
and low $x$ Dynamics}

\author{E. Elsen}

\address{DESY, Notkestr. 85,
D-22603 Hamburg, Germany\\ 
E-mail: Eckhard.Elsen@desy.de}


\maketitle
\abstracts{An attempt is made to illustrate the relation between low $x$ processes and diffraction. 
$ep$ scattering provides a unique laboratory, a single hadronic target probed by a point like lepton,  
where one can try to understand diffraction in terms of  a colourless exchange in QCD. Low $x$ processes 
eventually involve aspects of QCD which cannot be described perturbatively. The HERA inclusive measurements 
are examined in this respect and compared to results of the Tevatron.}

\section{Introduction}
The geometrical picture of diffraction is one in which the target is (essentially) left intact. The 
forward scattering amplitude is exponentially suppressed with momentum transfer squared  $t$. 
The slope $b$ of the distribution, $e^{bt}$ is directly related to the size of the scattering target. 

\begin{figure}[htb]
\begin{center}
\epsfxsize=0.4\textwidth
\epsfbox{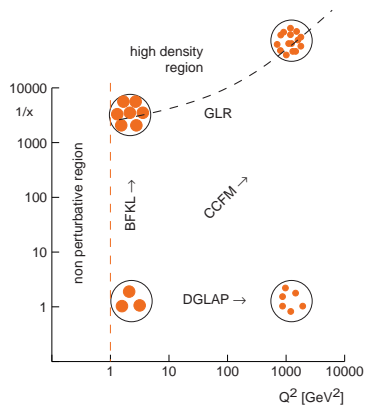}
\caption{The kinematic plane in $ep$ scattering and an illustration of the parton densities implied and 
the theoretical approximations deemed to be applicable.\label{fig:densities}}
\end{center}
\end{figure}

In $ep$ (and $pp$) scattering diffraction inevitably is also related to low $x$ such that the 
inelasticity $y$ remains $\approx 1$. Low $x$ naturally implies high mass $W$ since in $ep$ 
scattering $W^2=Q^2(1-x)/x$. Parton densities at low $x$ have been seen to be large\cite{ref:lowxincrease}. 
For large $Q^2$ there is hence a region of phase space which is calculable perturbatively in QCD even at 
low $x$. Figure~\ref{fig:densities} shows the regions of phase space in a pictorial manner. For very small 
$x$ initial states of high partonic density are formed. While in these regions methods of perturbative 
expansion\cite{ref:DGLAP,ref:BFKL} have to be proven to be valid it is interesting in its own right to 
search for experimental evidence for such new partonic configurations. In particular the relation between 
low $x$ processes and diffraction comes into play.

In diffraction the perturbative exchange has to be colourless and involves at least two gluons. 
Theoretically the interaction is thus modelled phenomenologically. A particularly useful description 
is placed in the proton rest frame. At low $x$ the $\gamma$ emitted from the incoming electron may 
turn into a $q\bar{q}$-dipole long before the interaction with the proton. The associated formation 
time $c\tau\sim 1/xm_p$ is typically much larger than the size of the proton target.

Various configurations of the $q\bar{q}$ states may be examined. In \emph{vector meson} production the 
pair forms a bound state that leaves the interaction unchanged. In \emph{deeply virtual compton scattering} 
(DVCS)\cite{ref:dvcs} the $q\bar{q}$ state even recombines to form a real photon. The study of \emph{inclusive 
diffraction} enables a test of these dipole configurations under controlled experimental conditions.

Vector meson production has been studied in much detail and a wealth of data both from HERA and the Tevatron 
have been quantitatively examined in terms of QCD.

\section{Geometrical Picture}
From its origin in optics diffraction is intimately associated with a geometrical 
interpretation\cite{ref:bartelsconf} - the size of the target which participates as a whole. 
The $t$-distribution is exponentially suppressed. The energy dependence arises when exciting 
higher angular momentum states that become accessible with increased phase space. This 
behaviour is beautifully supported by the observed energy dependence of the total $\gamma p$-cross section. 

\begin{figure}[ht]
\begin{center}
\epsfxsize=0.5\textwidth
\epsfbox{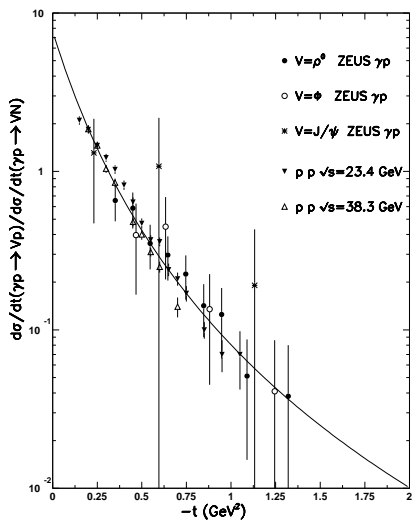}
\caption{The ratio of elastic to dissociative vector meson production as a function of the momentum 
transfer $t$ at the proton vertex for $\gamma p$ and $pp$-data.\label{fig:VMratio}}
\end{center}
\end{figure}

Vector mesons exhibit quite distinct variations in their $t$-behaviour reminiscent of the different sizes 
at work: from spatially extended $\rho$ mesons to much smaller $J/\Psi$ mesons. It is however interesting 
to compare the production of vector mesons in elastic and dissociative processes. While the elastic process 
is only affected by the formation of a vector meson, the dissociative process is modified by the form factor 
of the proton. Their ratio should hence be independent of the vector meson species. Figure~\ref{fig:VMratio} 
depicts the ratio as a function of the momentum transfer $t$ at the proton vertex, which seems to be described 
by a universal behaviour in support of this picture.

\section{Inclusive $ep$ Scattering}
So far the probe has been regarded as a quasi-real photon (without restrictions on transverse dimensions). 
For $Q^2\gg 0$ the probe becomes essentially point like and the spatial interplay between probe and target 
has to be considered. The transition between real photoproduction and deeply inelastic scattering is 
therefore particularly interesting.

\subsection{Inclusive $ep$ Scattering and the low $x$ Behaviour}
For $Q^2$ well below the $Z$-boson threshold ($Q^2<M_Z^2$) the cross section
\begin{equation}
\frac{{\rm d}\sigma}{{\rm d}x{\rm d}Q^2}
=\left(\frac{2\pi\alpha^2}{Q^4x}(1+(1-y))^2F_2-y^2F_L\right)
\end{equation}
\begin{figure}[htbp]
\begin{center}
\epsfxsize=\textwidth
\epsfbox{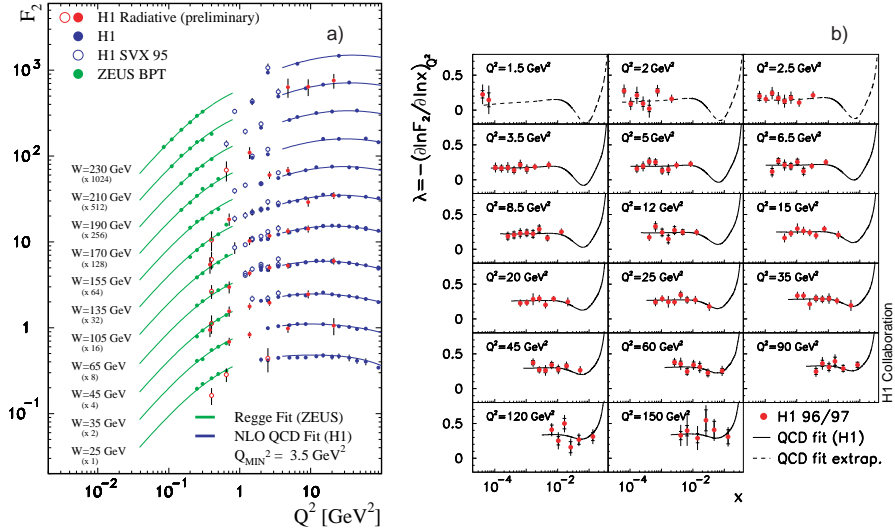}
\caption{a) The approach to small $Q^2$ of the structure function $F_2$. b) The $x$-derivative of 
the $F_2$-structure function vs $x$. The values $\lambda$ describe the low $x$ behaviour of the 
structure function according to $x^{-\lambda}$.\label{fig:smallqf}}
\end{center}
\end{figure}
is dominated by the structure function $F_2$; $F_L$ becoming relevant only at large $y$ and at small $Q^2$. 
Figure~\ref{fig:smallqf}a) shows the transition of $F_2$ towards photoproduction where $F_2$ has to vanish 
$\propto Q^2$ to yield a finite $\gamma^\ast p$ cross section. While the region $Q^2>1$\,GeV$^2$ is well 
described by recent QCD fits, the region below $1$\,GeV$^2$ relies on Regge inspired fits typically involving 
vector meson formation. For the region explored the energy dependence is smooth. There are no explicit signs 
of alteration of the behaviour as a function of $x\sim Q^2/W^2$.

It is therefore interesting to study the small $x$ behaviour of the structure function for not so small $Q^2$, 
i.e. the region related to the high energy behaviour of the scattering process and accessible to perturbative 
methods. Figure~\ref{fig:smallqf}b) depicts the derivative
\begin{equation}
\lambda = -\left(\frac{\partial \ln F_2}{\partial \ln x}\right)_{Q^2}\, .
\end{equation}
\begin{figure}[htbp]
\begin{center}
\epsfxsize=\textwidth
\epsfbox{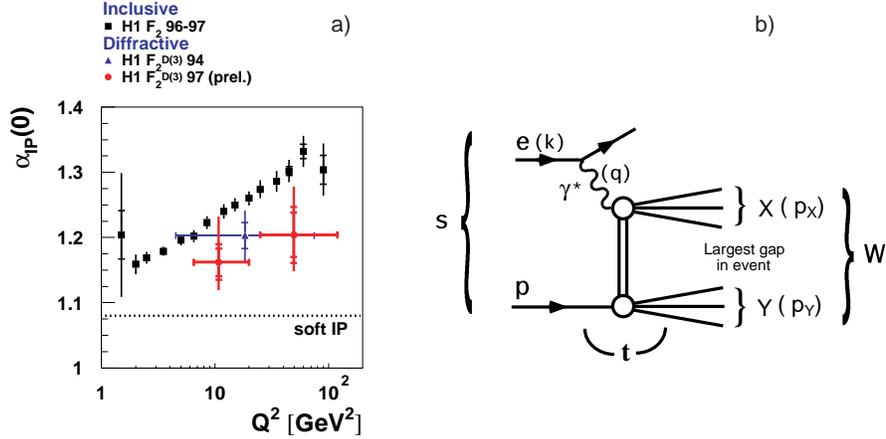}
\end{center}
\caption{a) the $Q^2$-dependence of the Pomeron intercept $\alpha_{\PO}(0)=1+\lambda$. Shown in red 
is the value derived from inclusive diffractive scattering, cf. section~\ref{sec:incldiff}. The 
diagram b) illustrates the definition of the signature for selection of diffractive events based 
on the signature of a large gap in rapidity.\label{fig:lambda}}
\end{figure}
The distribution is experimentally consistent with being constant for $x<0.01$ over a large range 
in $Q^2$. The implication $F_2=c(Q^2)x^{-\lambda}$ then yields a value for $\lambda$ that rises 
linearly with $Q^2$ (cf. figure~\ref{fig:lambda}a)) quite different from the photoproduction value 
$\lambda=0.08$. In the Regge language of \emph{Pomeron} exchange the exchange becomes harder the larger $Q^2$.

\subsection{Inclusive diffraction}
\label{sec:incldiff}
\begin{figure}[htbp]
\begin{center}
\epsfxsize=0.6\textwidth
\epsfbox{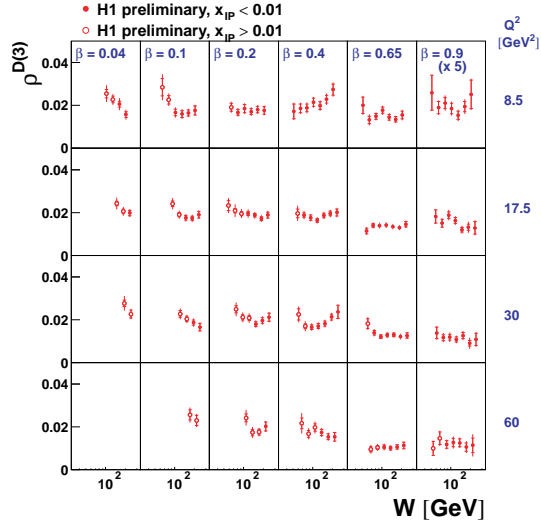}
\caption{The ratio $\rho^{D(3)}$ of the diffractive to inclusive structure function for two ranges 
of $\xpom$ from \cite{ref:H1diff}.  \label{fig:ratio}}
\end{center}
\end{figure}

Characteristic of the diffractive interaction is the large region void of hadronic activity - a gap in 
rapidity between the forward going nucleon and the centrally detected final state $X$, cf. 
figure~\ref{fig:lambda}b). Such a selection procedure leads to an inclusive measurement. 
Figure~\ref{fig:ratio} shows the fraction $\rho^{D(3)}(\beta, Q^2,\xpom)=M_X^2x/Q^2\cdot 
F_2^{D(3)}(\beta,Q^2,\xpom)/F_2(x,Q^2)$ of diffractively produced events as a function of 
$Q^2$ and $x$. With $\beta\approx{Q^2}/(Q^2+M_X^2)$ and for two intervals of 
$\xpom\approx{(Q^2+M_X^2)}/{(Q^2+W^2)}$ the fraction of diffractive events is considerable and 
the production persists up to the largest values of $Q^2$ of $60$\,GeV$^2$ investigated.
When analyzing the diffractive exchange in terms of a partonic description in QCD one finds a 
large gluonic component\cite{ref:H1diff}. To understand the formation of the final state more 
explicit final state measurements have to be investigated.

\subsection{Inclusive Diffraction of Charm Mesons and Jets}
\begin{figure}[htbp]
\begin{center}
\epsfxsize=\textwidth
\epsfbox{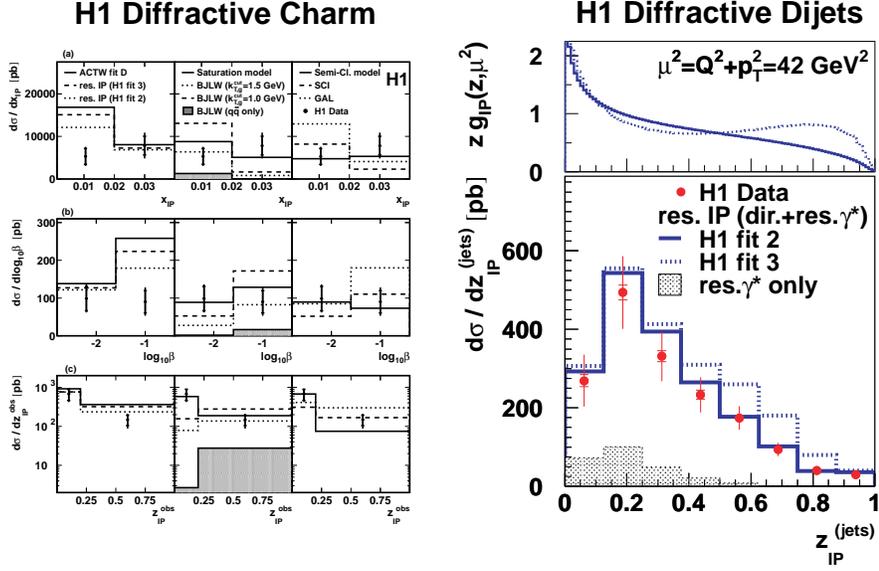}
\caption{Production of (left) charm and (right) jets in diffraction as a function of the momentum 
fraction $z_{\PO}$.  \label{fig:charmandjets}}
\end{center}
\end{figure}

Charm mesons originate from gluonic interactions and are hence a sensitive probe of the gluonic content 
of the interaction. Figure~\ref{fig:charmandjets} (left) demonstrates a large charm component in 
diffractive processes. Similarly, multi-jet formation involves several partons. From an analysis 
of such final states, cf. figure~\ref{fig:charmandjets} (right) one may conclude on the dynamical 
characteristics of the diffractive process. - As becomes evident, the description thus achieved is 
satisfactory and indicative of the large gluon content.

\section{Comparison to Tevatron Jets}
\begin{figure}[htbp]
\begin{center}
\epsfxsize=0.5\textwidth
\epsfbox{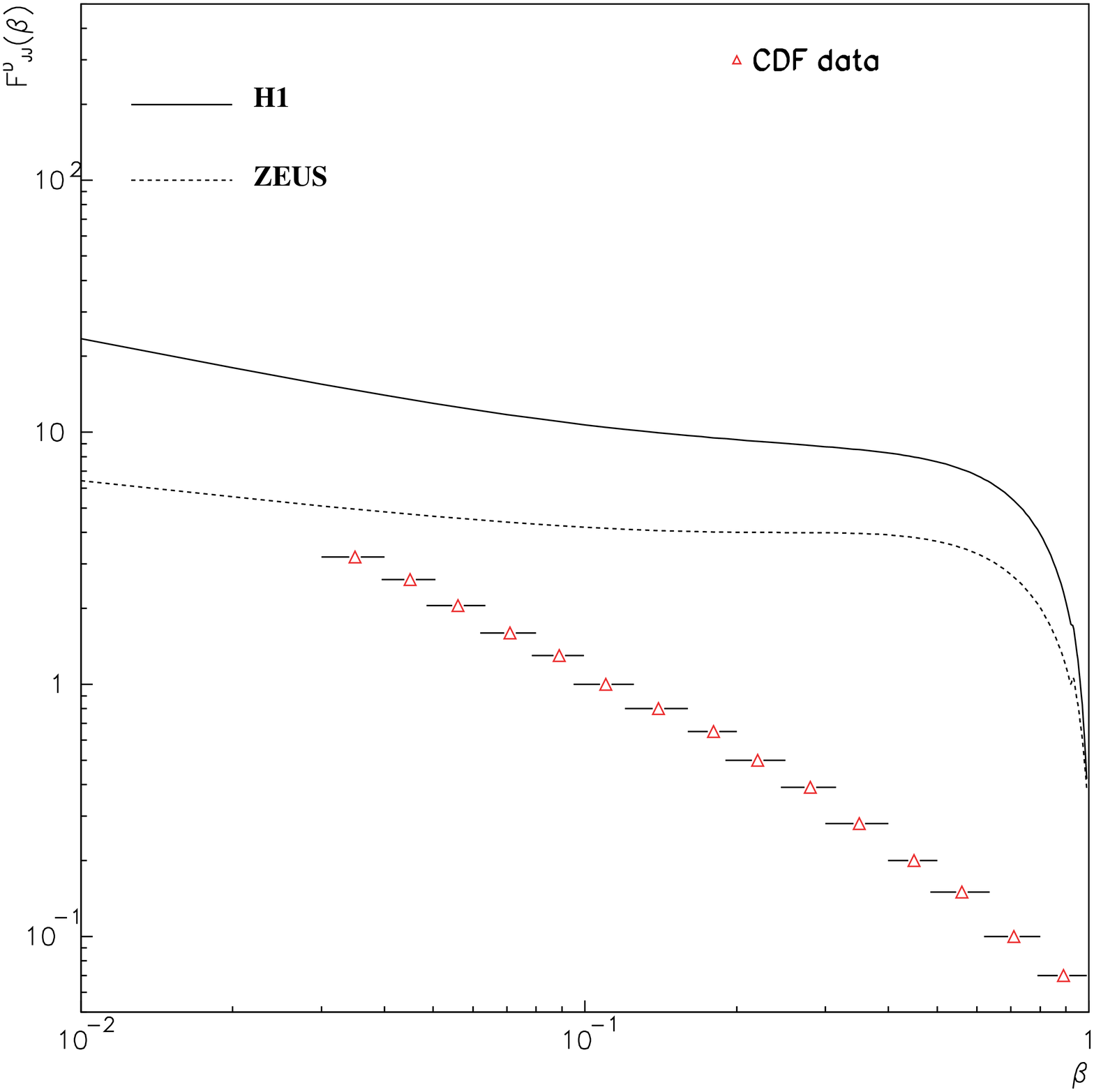}
\caption{Comparison of diffractive dijet production at the Tevatron with the expectations from 
parameterizations of the HERA measurements\cite{ref:bartelroyon}.  \label{fig:cdf}}
\end{center}
\end{figure}

The observations are quite at variance to the understanding of diffraction at the Tevatron. In $pp$ 
scattering the density of hadronic matter is so large that an initially formed gap is not likely to 
survive unaffectedly the remainder of the transition through hadronic matter. The application of the 
partonic description of diffractive $ep$ data thus leads to a gross overestimate of the production 
at the Tevatron, cf. figure~\ref{fig:cdf}. The environment of hadron-hadron scattering is less amenable 
to a theoretical understanding of diffraction.

\section{Conclusion}
As wealth of data has been amassed at HERA and the Tevatron on low $x$ physics and diffractive 
processes\cite{ref:savin,ref:milstead,ref:demortier}. The HERA data provide the clues on further 
understanding of high density (partonic) final states which to some extent are a prerequisite for 
diffractive interaction to take place. While at HERA diffractive production can be understood in 
terms of QCD the Tevatron data show that more complex mechanisms are at work in hadron-hadron 
scattering. The reach to low $x$ currently available at HERA does not yield experimental evidence 
for saturation of partonic matter density.

\section*{Acknowledgments}
I wish to thank organisers of the 31st ISMD for a very pleasant stay in Datong and the kind 
hospitality extended to the participants. Special thanks go to Liu Lianshou, Wu Yuanfang and 
Frans Verbeure for organizing the symposium and to Jim Crittenden for arranging a very fruitful 
session on diffraction.

\end{document}